\documentclass[sigconf]{acmart}

\usepackage{xcolor}

\AtBeginDocument{%
  }

\setcopyright{acmcopyright}
\copyrightyear{2018}
\acmYear{2018}
\acmDOI{XXXXXXX.XXXXXXX}

\acmJournal{POMACS}
\acmVolume{37}
\acmNumber{4}
\acmArticle{111}
\acmMonth{8}

\begin{document}

\title{Dynamic X-Ray Vision in Mixed Reality}

\author{Hung-Jui Guo}
\orcid{0000-0003-2233-846X}
\email{hxg190003@utdallas.edu}
\affiliation{%
  \institution{The University of Texas at Dallas}
  \streetaddress{800 W Campbell Rd}
  \city{Richardson}
  \state{Texas}
  \country{USA}
  \postcode{75080}
}

\author{Jonathan Z. Bakdash}
\orcid{0000-0002-1409-4779}
\email{jonathan.z.bakdash.civ@army.mil}
\author{Laura R. Marusich}
\orcid{0000-0002-3524-6110}
\email{laura.m.cooper20.civ@army.mil}
\affiliation{%
  \institution{U.S. DEVCOM Army Research Laboratory}
  \streetaddress{800 W Campbell Rd}
  \state{Texas}
  \country{USA}
  \postcode{75080}
}

\author{Balakrishnan Prabhakaran}
\orcid{0000-0003-0385-8662}
\email{bprabhakaran@utdallas.edu}
\affiliation{%
  \institution{The University of Texas at Dallas}
  \streetaddress{800 W Campbell Rd}
  \city{Richardson}
  \state{Texas}
  \country{USA}
  \postcode{75080}
}

\renewcommand{\shortauthors}{H.,J., Guo, J. Z. Bakdash, L. R. Marusich and B. Prabhakaran}

\begin{abstract}
X-ray vision, a technique that allows users to see through walls and other obstacles, is a popular technique for Augmented Reality (AR) and Mixed Reality (MR). 
In this paper, we demonstrate a dynamic X-ray vision window that is rendered in real-time based on the user's current position and changes with movement in the physical environment. Moreover, the location and transparency of the window are also dynamically rendered based on the user's eye gaze. We build this X-ray vision window for a current state-of-the-art MR Head-Mounted Device (HMD) -- HoloLens 2 \cite{hololensWeb} by integrating several different features: scene understanding, eye tracking, and clipping primitive. 
\end{abstract}



\begin{teaserfigure}
\begin{center}
  \includegraphics[width=0.9\textwidth, keepaspectratio=true]{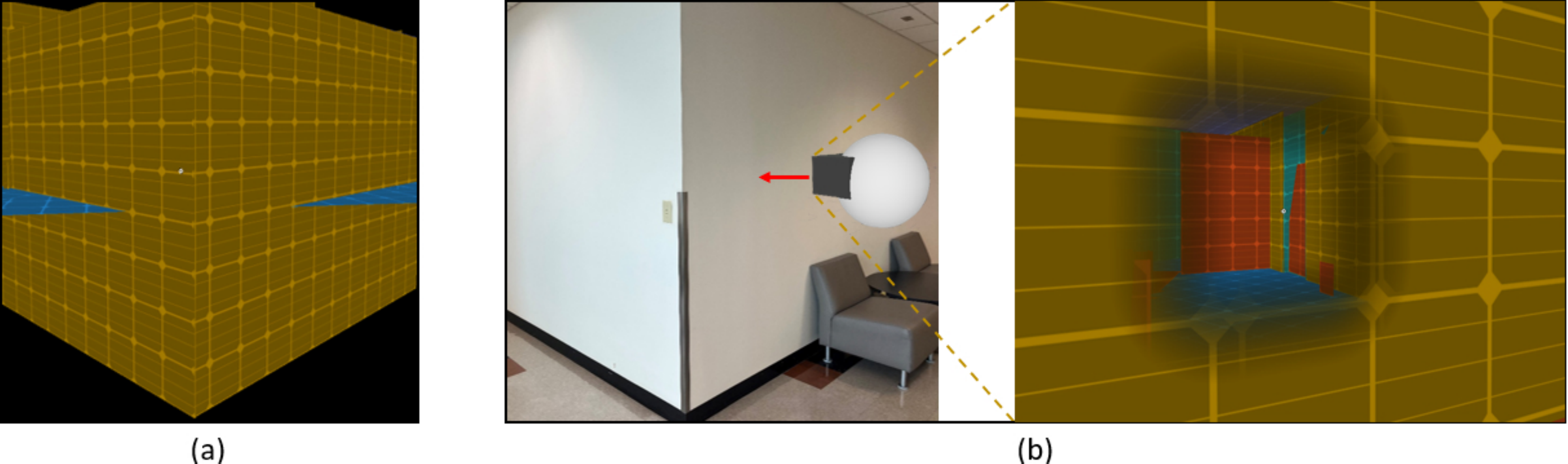}
  \caption{(a) Side view of the created collaborative scene understanding generated augmented reality environment based on an indoor room, (b) Dynamic X-ray vision window used by a user to navigate the indoor environment}
  \label{fig:Teaser}
\end{center}  
\end{teaserfigure}

\maketitle

\section{Introduction}
Generally, in a real environment it is almost impossible for a user to obtain information occluded by environmental objects, such as a chair behind a wall, when the user has no prior understanding of the environment. 
Even when advanced information about the structure of the environment is available and presented with a tool such as a minimap, physical occlusions still challenge users' ability to obtain direct visual information about objects and the spatial layout of the beyond-line-of-sight environment.

A virtual X-ray vision window allows a user to directly view a target object through a physical occlusion at a fixed point, enabling quick and accurate judgments about the occluded target object while not sacrificing knowledge about the occluding surrounding environment. By providing information otherwise unavailable in the real-world environment, X-ray vision has potential law enforcement and military applications (e.g., knowledge of the locations of beyond-line-of-sight threats and civilians prior to entering a room) \cite{Phillips2021}.  
Examples of previous implementations of X-ray vision systems include robot with a stereo depth camera \cite{Phillips2021} and image-based rendering \cite{Avery2009}. In both of these implementations, the X-ray window is a flat image aligned with a physical surface.

In contrast to previous approaches, virtual content in our X-ray window is rendered dynamically using scene understanding. That is, the virtual window updates in real-time based upon the user's position, movement, and eye-gaze in the physical environment. Consequently, our X-ray vision system mimics a physical window. It provides non-pictorial depth cues such as motion parallax (changes in the perspectives of stationary objects with self-motion) and binocular disparity (slightly different visual images in each eye due to their horizontal separation) typically available in real-world environments \cite{cutting1995perceiving}. This allows users to move and still naturally see through physical occlusions (such as real walls, created by the scene understanding observer) as if looking through a window in the real-world. 

\section{Dynamic X-Ray Vision Window}
The dynamic X-ray vision window is built in the HoloLens 2 device \cite{hololensWeb}, a Head-Mounted Device (HMD) that includes the Mixed Reality (MR) environment. This system incorporated the virtual environment created by HoloLens 2 Scene Understanding feature pre-built in the Mixed Reality ToolKit (MRTK) (\url{https://docs.microsoft.com/en-us/windows/mixed-reality/mrtk-unity/mrtk2/}). We use the long-throw depth camera to observe the surrounding real-world environment; we call this observed environment a collaborative scene understanding generated augmented reality environment, see Figure \ref{fig:Teaser} (a). 

In addition, to further help users see through the occluding real-world surface from a distance, the X-ray vision window uses the clipping primitive feature created by the MRTK. This makes the contact area of the touched scene objects partially transparent, as shown in Figure \ref{fig:Teaser} (b). Moreover, we also use the gaze-tracking feature in HoloLens 2 to obtain real-time eye direction/movements and its intersection with scene objects 
to determine the location of the X-ray vision window.


Using a dynamic window for X-ray vision based on user position/movement and eye-gaze is an advancement over image based X-ray vision. This X-ray vision has the potential to provide benefits over prior implementations. However, the usability and human performance with our system needs to be evaluated. For example, it is possible users may find that interacting with the X-ray vision window is unfamiliar, uncomfortable, or even distracting. Using previous behavioral research on X-ray vision in AR/MR as a reference (see \cite{Livingston2013} for a summary), we propose a user evaluation of the dynamic X-ray vision window demonstrated here.

\subsection{Proposed User Study Design}
Among previous evaluations of X-ray vision in AR/MR, the most common method is to assess how an X-ray vision window may affect participants' depth perception of physical/virtual objects (\cite{Phillips2021, Zollmann2014}).
Other methods include testing whether the participants can understand occlusion layers \cite{Kitajima2015}, or asking the participants to acquire target objects under different X-ray vision settings \cite{Sandor2010}.

According to \cite{Livingston2013}, most previous X-ray vision studies are conducted while participants are stationary. In our proposed study, We will incorporate participants' physical movement, allowing them to explore the environment, while dynamically updating virtual information in real-time. This feature allows us to explore the benefit of our X-ray vision window system and design a more realistic user study with visual depth cues that are not present when stationary (e.g., motion parallax).

Therefore, we plan to build a game-like user study to explore how a dynamic X-ray vision window (as compared to other forms of X-ray vision or other types of advanced spatial knowledge) may benefit human performance, measured by physical movement direction and/or reaction time. The study will include target objects placed behind occluders in a physical environment (such as an indoor room), and will require participants to move in the environment and acquire target objects with and without the X-ray vision window. 

\section{Conclusion and Future Work}
In this paper, we present a new dynamic X-ray vision window built in HoloLens 2, which can render beyond-line-of-sight content in real-time based on user position and movement with eye-gaze direction to potentially improve usability. In the future, we will evaluate and improve this proposed system according to the user study design described in the previous section and make this system more realistic and close to the physical world by attaching textures to virtual surfaces.

\section*{Acknowledgments}
This research was sponsored by the DEVCOM U.S. Army Research Laboratory under Cooperative Agreement Number W911NF-21-2-0145 to B.P. \\
The views and conclusions contained in this document are those of the authors and should not be interpreted as representing the official policies, either expressed or implied, of the DEVCOM Army Research Laboratory or the U.S. Government. The U.S. Government is authorized to reproduce and distribute reprints for Government purposes notwithstanding any copyright notation.

\bibliographystyle{ACM-Reference-Format}
\bibliography{X-rayVisionMain}

\end{document}